\pdfoutput=1

\documentclass[aps,prl,twocolumn,superscriptaddress,floatfix]{revtex4}

\usepackage{graphicx}

\begin{document}

\title{Topology of the polarization field in 
       ferroelectric nanowires from first principles}

\author{J. W. Hong}
\affiliation{Department of Earth Sciences, University of Cambridge, 
             Downing Street, Cambridge CB2 3EQ, UK}
\affiliation{AML Department of Engineering Mechanics, Tsinghua University, 
             Beijing, 100084, P. R. China}
\author{G. Catalan}
\affiliation{Department of Earth Sciences, University of Cambridge, 
             Downing Street, Cambridge CB2 3EQ, UK}
\author{D. N. Fang}
\affiliation{AML Department of Engineering Mechanics, Tsinghua University, 
             Beijing, 100084, P. R. China}
\author{Emilio Artacho}
\affiliation{Department of Earth Sciences, University of Cambridge, 
             Downing Street, Cambridge CB2 3EQ, UK}
\author{J. F. Scott}
\affiliation{Department of Earth Sciences, University of Cambridge, 
             Downing Street, Cambridge CB2 3EQ, UK}
\affiliation{Cavendish Laboratory, University of Cambridge, 
             JJ Thomson Ave, Cambridge CB3 0HE, UK}

\date{\today}

\begin{abstract}
  The behaviour of the cross-sectional polarization field is explored for 
thin nanowires of barium titanate from first-principles calculations.
  Topological defects of different winding numbers have been obtained, 
beyond the known textures in ferroelectric nanostructures.
  They result from the inward accommodation of the polarization
patterns imposed at the surface of the wire by surface and edge effects.
  Close to a topological defect the polarization field orients out of 
the basal plane in some cases, maintaining a close to constant magnitude, 
whereas it virtually vanishes in other cases.
\end{abstract}


\maketitle


  The drive towards ever smaller ferroelectric devices~\cite{Scott06} 
has resulted in novel geometries such as nanowires~\cite{Urban02,Yun02}, 
nanotubes~\cite{Luo03,Morrison03,Mao03}, and nanodots~\cite{Shin05,Chu04}.
  It was initially assumed that as the size of such nano-shapes continued 
to decrease, the ferroelectric polarization would vanish: ferroelectricity 
is a collective phenomenon after all.
  However, seminal works by Fu, Bellaiche and Naumov~\cite{Fu03,Naumov04} 
suggested that, before vanishing, the polarization may first 
undergo a size-induced transition by which the dipoles form a ferroelectric 
vortex, a polar configuration that does not exist in bulk and persists 
down to very small diameters.
  This insight has triggered much work, both theoretical~\cite{Geneste06,
Morozovska06,Prosandeev08,Hong08,Wang08-2,Shimada09} and 
experimental~\cite{Spanier06,Wang06,Rodriguez09} towards establishing true 
critical diameters for ferroelectricity, understanding the characteristics of 
ferroelectric vortices, and establishing their existence.

  In the present work we have used ab-initio methods to calculate the polar 
configuration in ultra-thin nanowires of the archetypal ferroelectric 
BaTiO$_3$ (BTO).
  Our results show that the topological landscape of the polarization 
field $\vec P(\vec r)$ in nanowires is much more complex than hitherto 
assumed, with polar configurations other than vortices appearing.
  The new topological textures include saddle points and quadrupoles as well 
as more complex configurations, all of them well described within the classic 
framework of topological defects of different winding numbers~\cite{Mermin79}.
  Such patterns are induced by the surfaces and can be understood 
in terms of simple surface and edge effects.
  Of equal importance is the discovery that the axial component of the 
polarization can either decrease or increase towards the surface of the wire 
depending on its termination, thus resolving the long-standing controversy on
the sign of the so-called extrapolation length~\cite{Kretschmer79,Zhong94,
Musleh09}.
  More importantly, the critical size for the persistence of polarization in 
nanowires is found to be determined by their surfaces.

  The two-dimensional (2D) topology of of the polarization field in thin 
nanostructures can be mapped onto the topology of $(P_x , P_y)$ for wires 
if the field remains constant along $z$, the axial direction of the wire. 
  This paper focuses on topological defects in such 2D field~\cite{Mermin79}
(these so-called ``defects" do {\it not} refer to structural defects of the 
wires).
  The winding number is defined as $n=\phi / 2\pi$, being $\phi$ the total 
angle the 2D field vector rotates when going around a closed circuit. 
  For a continuous field $n$ must be an integer, which if different from
zero implies the existence of at least one defect within the circuit.
  Any smooth deformation of a vector field will conserve $n$, while changing 
the winding number implies a dramatic, extended and highly energetic 
rearrangement
\footnote{Strictly speaking, the topological defects mentioned here arise for
vector fields of constant magnitude, giving rise to a discontinuity in the 
field derivatives at the defect.
  Ferroelectric fields (or their proxies, e.g. cation off-centerings)
are of variable magnitude and vanish at the  defect.}.
  Such $\Delta n \neq 0$ rearrangements have been used to define ``phase 
transformations"~\cite{Naumov07} for nano-objects displaying vortices ($n=+1$). 
  Here we find realizations of$n=0,1,-1$, and $-3$.
  We see textures that were described before (e.g. radial~\cite{Hong08-2}) 
together with completely novel ones. 
  The different textures are induced (and thus can be manipulated) by the
different surface terminations of the wires.

\begin{figure}[t]
\begin{center}
\vspace{-1mm}
\begin{tabular}{cc}
\includegraphics[width=0.22\textwidth]{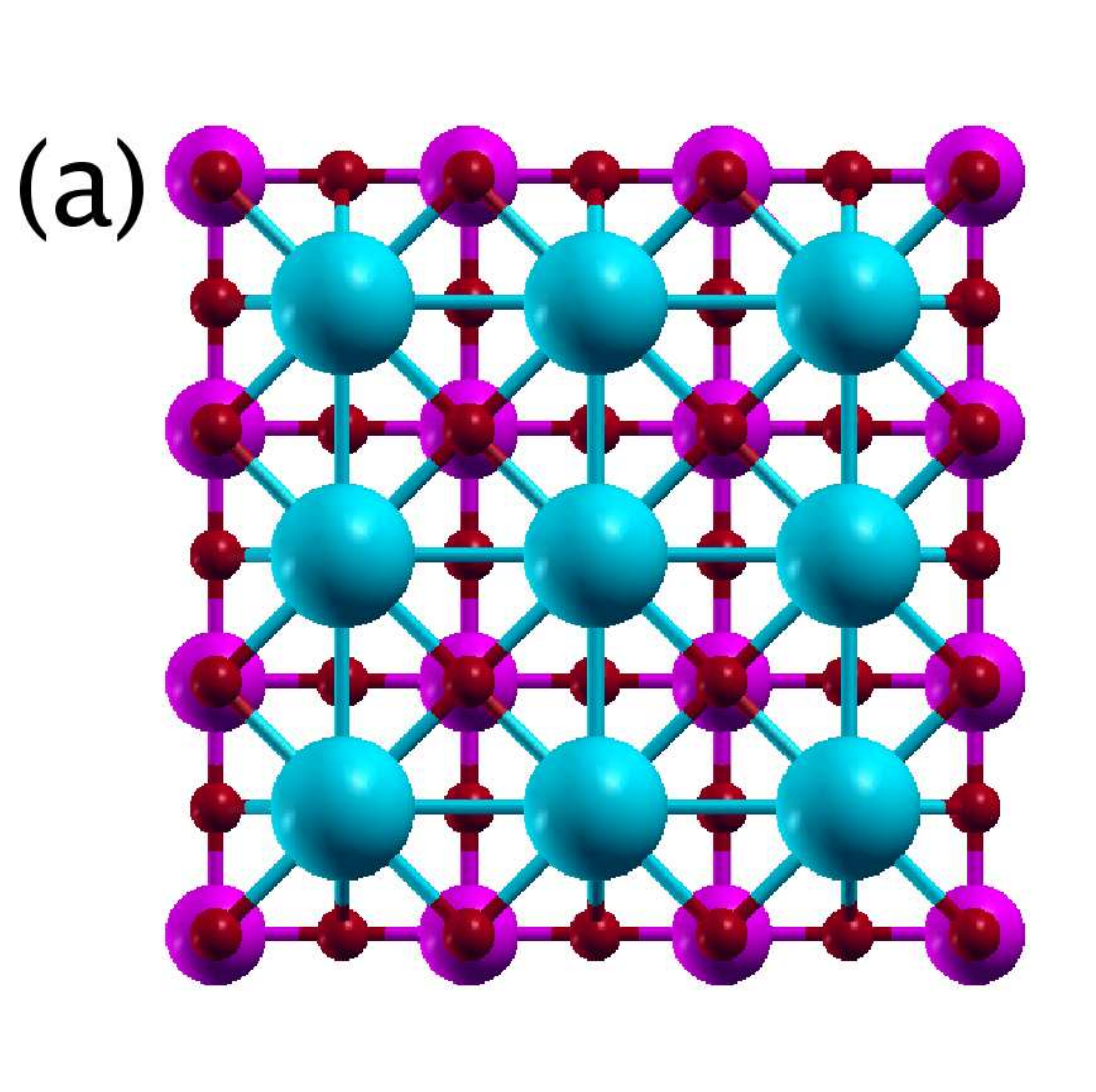}
\includegraphics[width=0.22\textwidth]{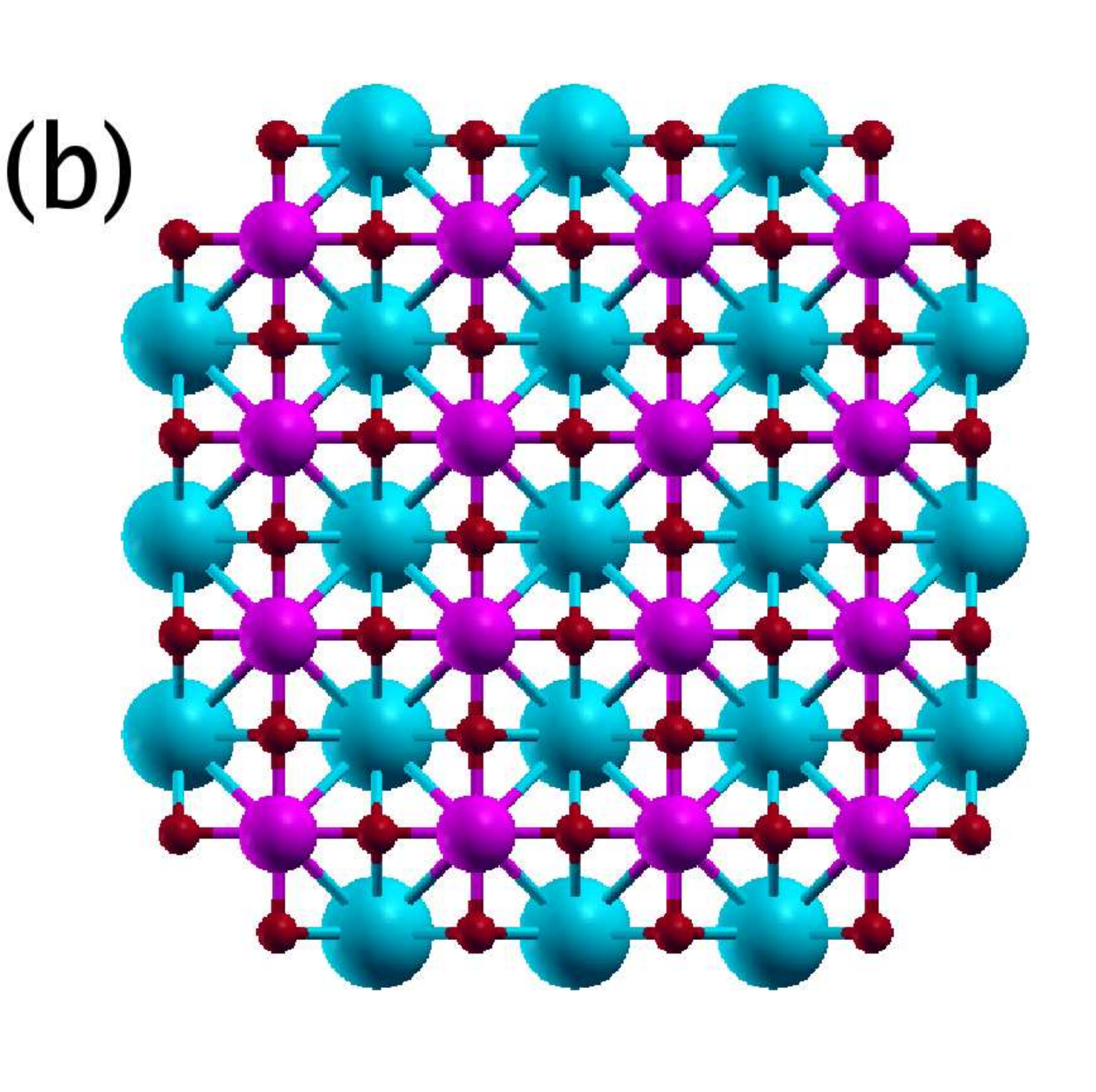} \\
\includegraphics[width=0.22\textwidth]{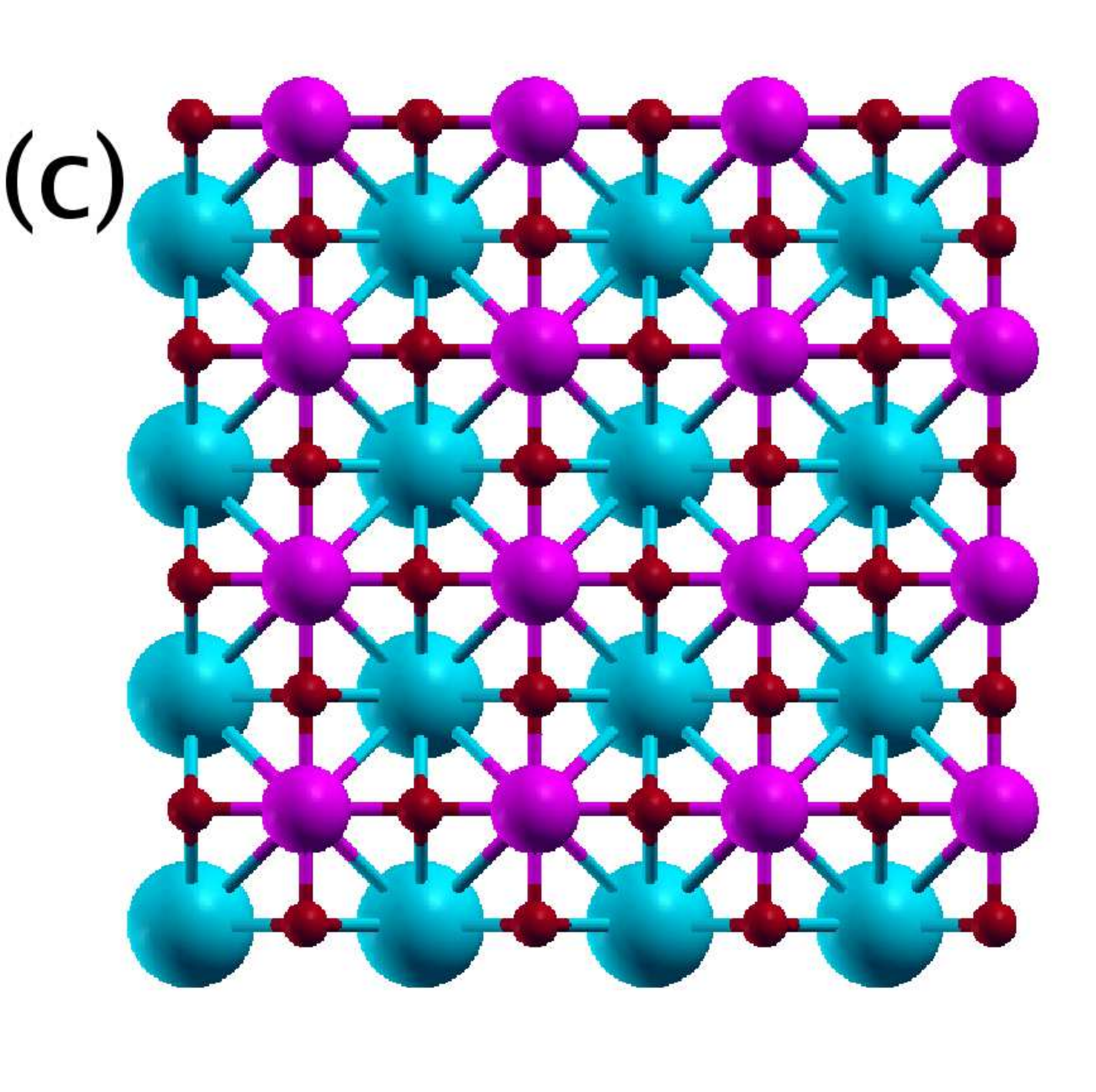}
\includegraphics[width=0.22\textwidth]{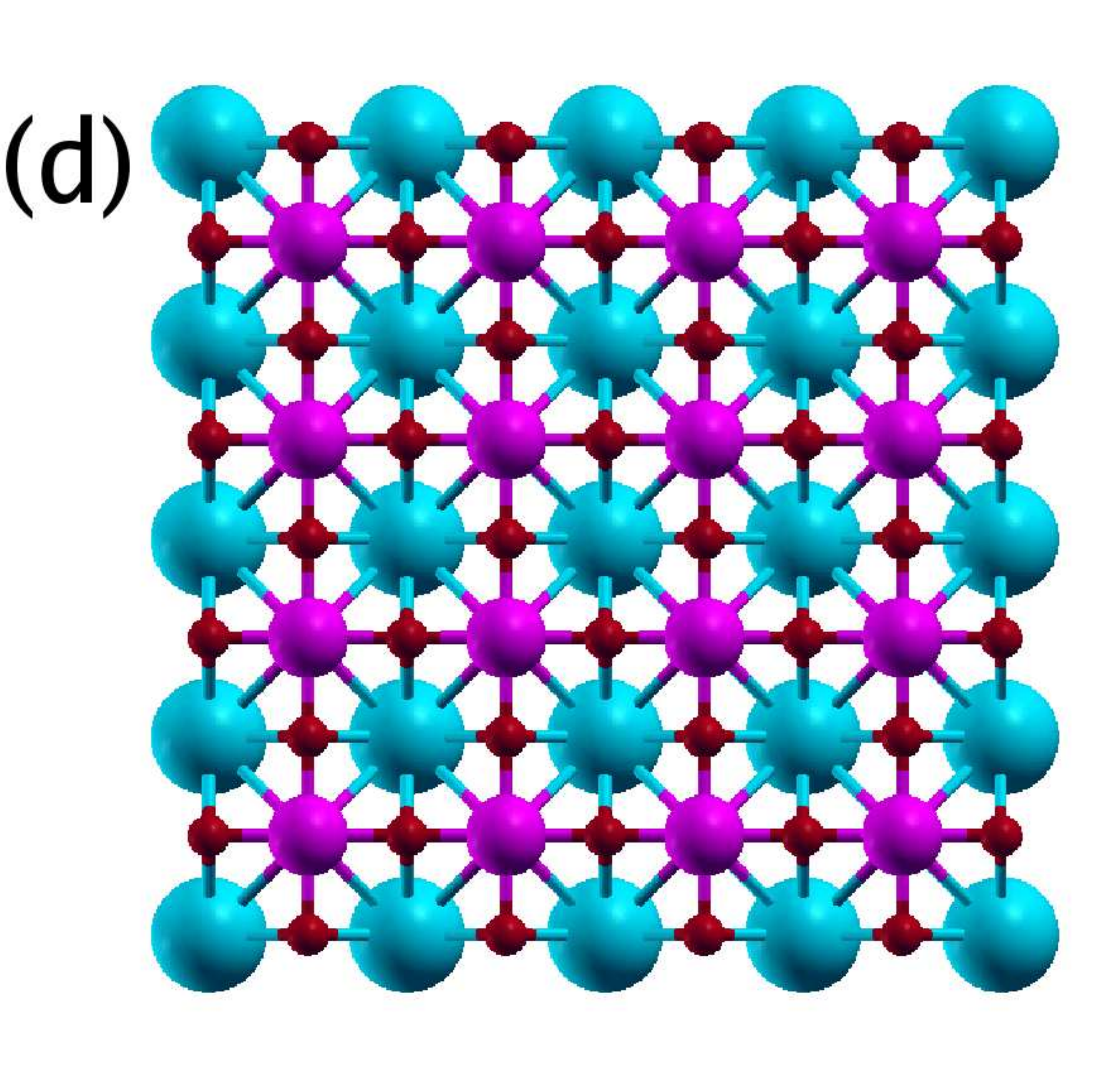}
\end{tabular}
\vspace{-1mm}
\caption{\label{SECT}{(Color online) 
  Sections of the different BaTiO$_3$ nanowires considered.
  The largest balls (cyan) are for Ba ions, the intermediate ones (magenta) 
represent Ti ions, and the smallest ones (red) are for O. 
  (a) is TiO$_2$ terminated,  
  (b) is BaO terminated, except the edges, where lines of Ba atoms have been
removed,
  (c) is stoichiometric, with two BaO surfaces and two TiO$_2$ ones.
  (d) is BaO terminated.}}
\end{center}
\end{figure}


  Our calculations were performed within DFT and the generalized gradient 
approximation (GGA)~\cite{Wu06}.
  We used the {\sc Siesta} method~\cite{Ordejon96,Soler02}, based on 
finite-range numerical atomic orbitals, using a double-$\zeta$ polarized basis 
set~\cite{Anglada02,Geneste06}.
  Norm-conserving pseudopotentials were used, including into the valence the 
semicore shells 3$s$ and 3$p$ for Ti and 5$s$ and 5$p$ for Ba.
  The performance of the method was tested for bulk BaTiO$_3$.
  The obtained lattice parameter for the rhombohedral phase is 0.12 \% smaller
than the experimental value of 4.0036 \AA\ at 15~K~\cite{Kwei93}, and
0.055 \% smaller than the published results with the same GGA and plane
waves as basis set~\cite{Wu06}.
  The rhombohedral angle was found to be 89.84$^{\rm o}$, exactly as obtained
in the same experiments, and the off-centering of the Ti atom came out as 
0.184 \AA, as compared with the experimental 0.185 \AA.
  Periodic boundary conditions were used with a simulation cell describing 
a square array of infinite nanowires, such that periodic replicas are 
separated by 24 \AA.
  The width of the wires is shown in Fig.~1.
  The unit cell along the wire axis comprises a single bilayer in all 
cases, and therefore all polarization textures are homogeneous
along $z$ by construction.
  Integrals in reciprocal space used a $\vec k$-mesh of 10 \AA\ 
cutoff~\cite{Moreno92}, while the integrals in real space used a
$\vec r$-mesh of 350 Ry cutoff~\cite{Soler02}. 
  The atomic positions were relaxed until all atomic force components 
were smaller than 10 meV/\AA. 
  From the relaxed structures, the Ti displacement from the centre of 
its coordination octahedron was taken to define $\vec P (\vec r_{\rm Ti})$, 
as a measure of the local dipole.
  For Ti atoms on surfaces or edges the off-center displacement was 
determined by the remaining coordinating O atoms.
  It should be emphasized, however, that the focus of the paper is on the 
topological aspects of the polarization, which are not affected by quantitative 
details related to this definition.

  Four different kinds of nanowires have been considered, corresponding to
different surface terminations, as shown in Fig.~1, all showing (100)
and (010) surfaces, or slight alterations thereof.
  Two of them display the same termination on all sides, namely, all
TiO$_2$-terminated (a), or all BaO-terminated (d). 
  Wire-type (c) is stoichiometric, with both terminations, and corresponding
to the wires studied in ref.~\cite{Geneste06}.
  Finally (b) is like (d) except for the fact that the Ba on the edges has
been removed.
  Each wire type has been considered in several sizes, mainly $4 \times 4$,
and $5\times 5$, counting the Ti atoms.



  Fig.~2 shows the polarization textures of the wire types of Fig.~1,
alongside qualitative sketches of the corresponding topological defects for a 
2D vector field.
  The discrete set of off-centering vectors is assimilated to a 
lattice discretization of a 2D $\vec P$ field.
  Different wire widths correspond to different discretizations of the
corresponding general pattern.
  A first striking result is that even for the small sizes 
considered (and the coarse discretization they imply) the field textures
appear very clearly.
  All textures preserve inversion symmetry at
the center of the wire if it was not previously broken by the surface 
termination.
  The first two cases in Fig.~2, displaying inward and outward radial 
patterns, correspond to the same kind of topological defect, of winding
number 1. 
  They are thus homotopical with each other, but also with the vortex 
structures seen in refs.~\cite{Naumov04,Prosandeev08}.
  The continuous transformation taking a radial texture onto a vortex one
is beautifully illustrated in Mermin's review~\cite{Mermin79}.
  They of course differ in other physical aspects, such as the 
non-zero toroidal moment of the vortex patterns. 

\begin{figure}[t]
\begin{center}
\begin{tabular}{cc}
\includegraphics[width=0.21\textwidth]{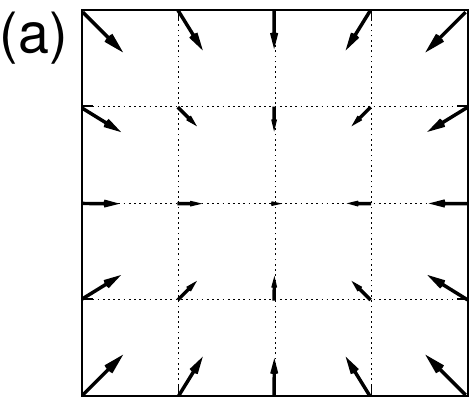}
\includegraphics[width=0.21\textwidth]{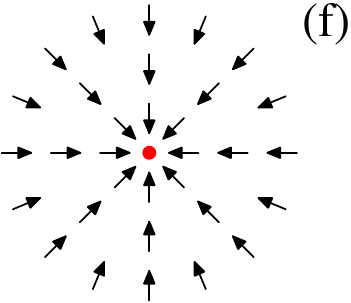} \\
\includegraphics[width=0.21\textwidth]{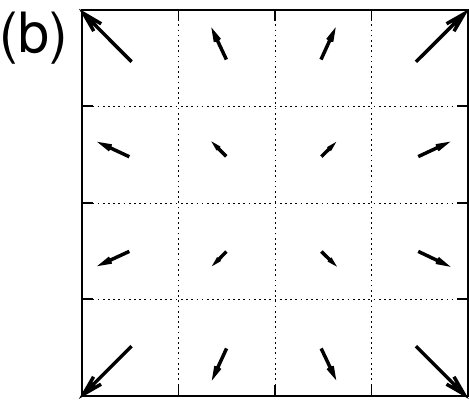}
\includegraphics[width=0.21\textwidth]{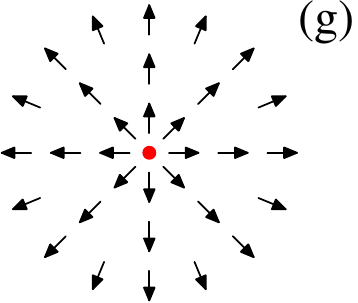} \\
\includegraphics[width=0.21\textwidth]{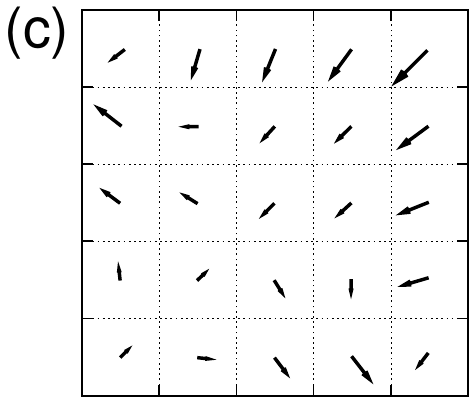}
\includegraphics[width=0.18\textwidth]{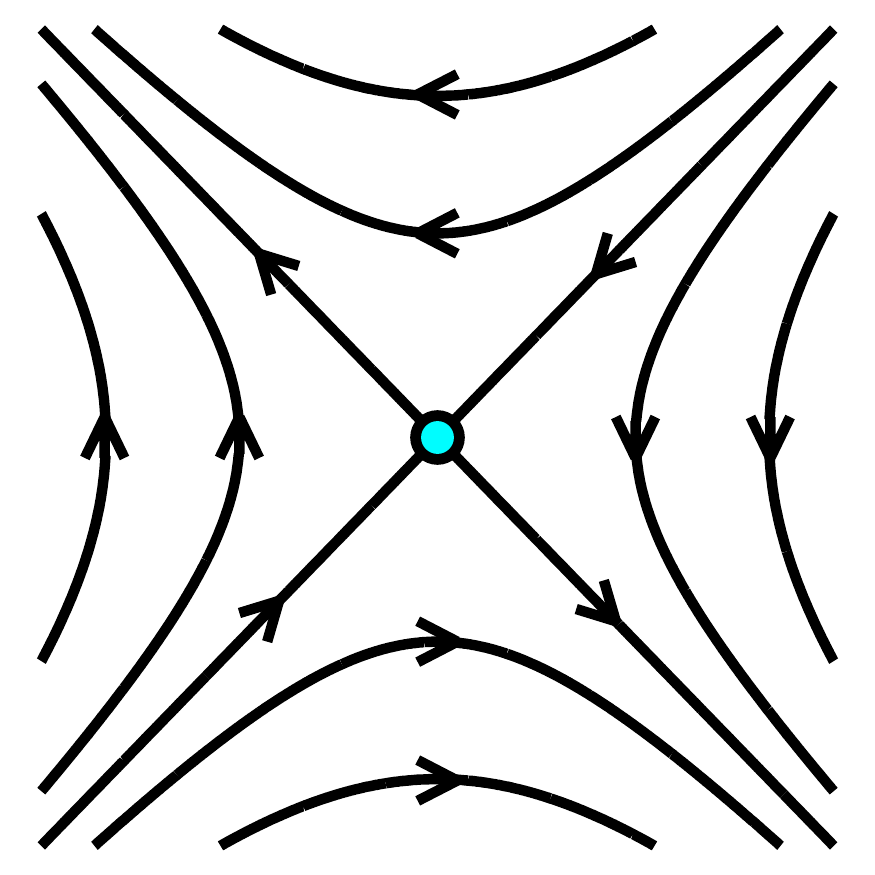} 
\includegraphics[width=0.035\textwidth]{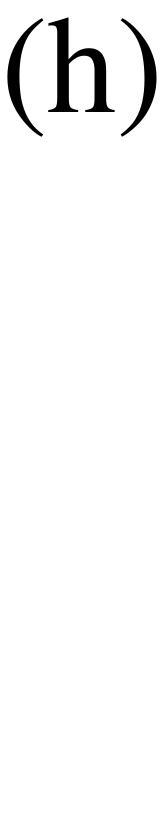} \\
\includegraphics[width=0.21\textwidth]{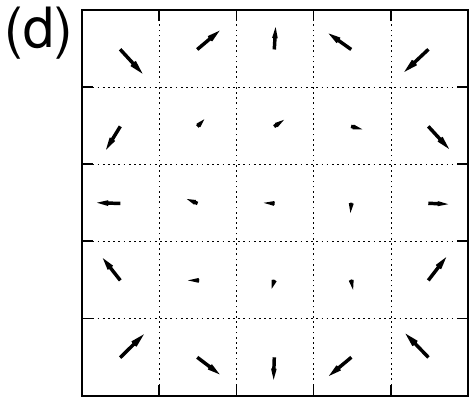}
\includegraphics[width=0.18\textwidth]{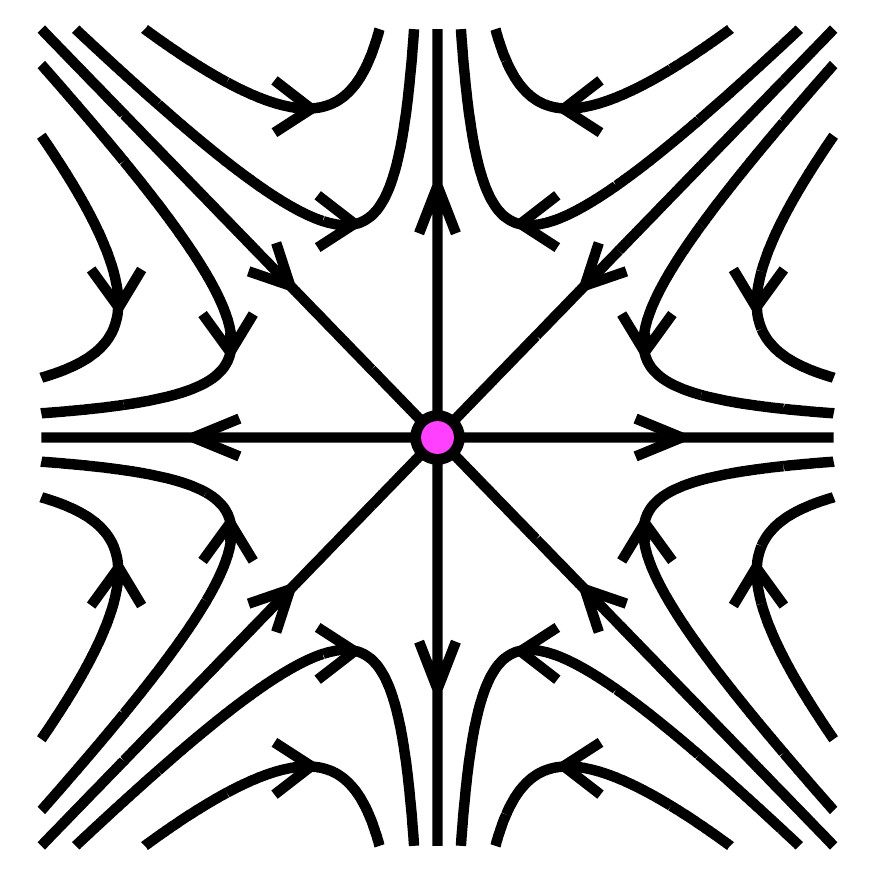}
\includegraphics[width=0.035\textwidth]{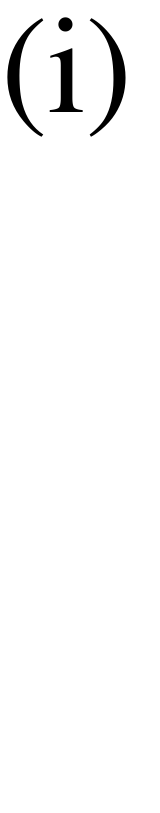} \\
\includegraphics[width=0.21\textwidth]{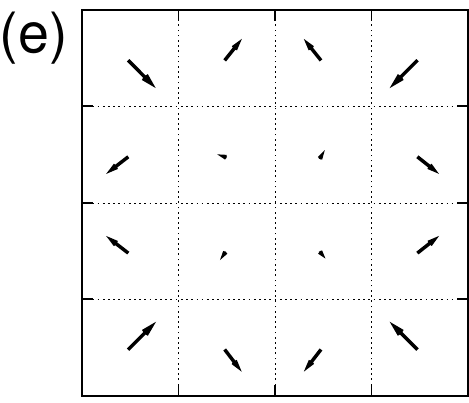}
\includegraphics[width=0.18\textwidth]{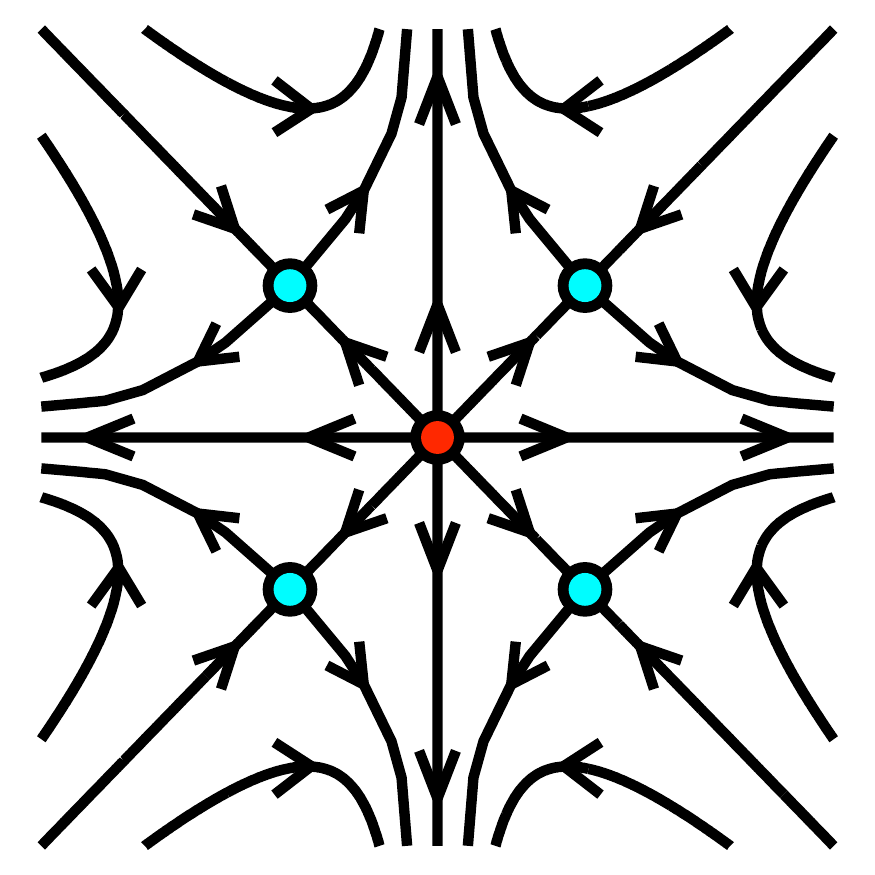}
\includegraphics[width=0.035\textwidth]{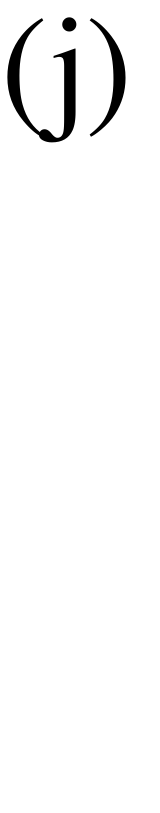}
\end{tabular}
\caption{\label{ZDISP}{Left column (a)-(d): Ti off-center displacements for 
the four respective nanowires in Fig.~1. (e) is analogous to (d) but for a 
thinner wire. The right column sketches the corresponding 
2D field lines around topological point defects of winding numbers 
$n=1,1,-1,$ and $-3$, respectively, (j)  showing the $n=-3$ 
decomposition into a central $n=+1$ and four $n=-1$ defects. 
The respective largest arrows corresponds to (a) 21 pm, (b) 35 pm, 
(c) 23 pm, (d) 6.8 pm, and (e) 6.6 pm. }}
\end{center}
\end{figure}

  Figs.~2(c) and (d) correspond to different winding numbers as yet not 
observed or proposed in this field, namely $-1$ and $-3$.
  In Fig.~2(c) the defect is displaced from the center (it is the only 
non-centrosymmetric wire).
  Fig.~2(d) shows the $n=-3$ case of the BaO terminated wire.
  The high winding number imposes a more rapidly varying field around
the defect, which implies higher energy.
  The system responds by a more pronounced depression of the magnitude of 
the 2D field around the defect.
  The winding of the field around the outer circuit in the wire section
clearly corresponds to $n=-3$.
  
  Fig.~2(e) is for the same BaO termination as Fig.~2(d) but  
smaller thickness.
  Although a different discretization, the winding around the outer circuit
clearly shows the same $n=-3$.
  In this case the inner displacements, although small, all point mainly 
outwards, inconsistently with the the $n=-3$ texture: This system has found 
more stability in decomposing the global $n=-3$ global defect into a set of 
defects whose winding numbers add up to $-3$.
  In general~\cite{Mermin79}, when following the winding of the field around
a circuit the winding number obtained is the sum of winding numbers of all
the point defects enclosed in the circuit.  
  In this case the texture can be described by a central defect with $n=+1$, 
the field radiating outward as in Fig.~2(g), plus four other 
defects of $n=-1$, as illustrated in the last sketch.
  

  The described textures are clearly induced by surface and edge effects.
  Indeed, the phenomenlogy described can be understood as originated by
three simple tendencies: 
  $(i)$ In Ti terminated surfaces and edges, the Ti atoms tend to off-center
inwards,
  $(ii)$ Ba surface termination induces outward Ti off-centering patterns,
while
  $(iii)$ Ba edge termination induces an inward tendency.
  The competition of $(ii)$ and $(iii)$ in the Ba terminated wires gives
rise to the rich $n=-3$ pattern. 
  By removing the Ba edges [Fig.~2(b)], the competition disappears and the 
fully outward pattern is established.
  Similarly, when removing the Ba edge in the stoichiometric case the 
polarization field becomes topologically homogeneous
($n=0$), with the field lines crossing the wire diagonally pointing left and 
down (not shown).
  Special chemical tendencies at edges and surfaces have been previously
reported~\cite{Spanier06,Shimada09}.
  Shimada {\it et al.}~\cite{Shimada09} talks specifically about strong
``edge bonds" for PbTiO$_3$ nanowires.


  The surfaces not only determine the $(P_x,P_y)$ topology, but also the 
$P_z$ behaviour (Fig.~3).
  Axial switchable polarization has been observed experimentally~\cite{Wang06},
and has been analyzed within Landau-Ginzburg theory~\cite{Morozovska06,Hong08},
and DFT~\cite{Geneste06,Shimada09}.
  We find that for TiO$_2$ termination the $4 \times 4$ wires display 
$P_z=0$, while it is finite for $5 \times 5$ (in the figure) or larger. 
  When non-zero, the axial polarization stems mainly from the surface cells, 
the central cells displaying much smaller displacements.
  This is surprising: an essentially constant-magnitude field is
commonly assumed for these and other ferroelectric systems (some models
have this assumption built in, as a constant dipole moment of arbitrary
orientation).
  Here, however, the field responds by vanishing smoothly into the defect.
  The BaO terminated wires respond in the more expected way: 
discontinuity at the defect is avoided without substantial diminishing of
the polarization field, by having maximal axial polarization at the wire
center.
  This is the case for both with and without Ba edges, although the latter
shows a much smaller axial component at the center.
  There is also a $3\times 3$ critical thickness for $P_z$ (largest system
with $P_z=0$) in the Ba terminated case.
  In the stoichiometric wires the situation is different.
  The figure shows a $5 \times 5$ case (the critical thickness for this type 
is $2 \times 2$, in agreement with ref.~\cite{Geneste06}), which again shows 
large $P_z$ for the surface and edge Ti displacements, for which 
the in plane components were also large.

  In addition to their instrinsic fundamental interest, the polarization-field 
textures and the possibility of manipulating them, offer interesting 
exploitation possibilities.
  Net polarization has been measured and switched perpendicularly to the wire 
axis at room temperature with a scanning probe 
microscope~\cite{Urban02,Yun02,Spanier06}. 
  Such a setup should thus respond as a three-state system, with +1, 0 and 
-1 states, the +1 and -1 corresponding to the conventional ferroelectric 
response, and the zero to the stable topologically defective state 
(with $|P|$ much smaller than for the other states, strictly zero if the 
unpolarized wire is centrosymmetric). 
  The field rearrangement needed for a non-homotopical change of texture 
would mean an energy barrier when changing the net polarization, giving a 
scenario analogous to a first order phase transition above the transition 
temperature: the global minimum for $P\sim 0$ and two local minima for + and -.
  Note that some of the $n\neq 0$ textures exhibit net quadrupoles.
  These configurations can be switched (destabilized) by application of an 
electric field gradient, which could in principle be used for a quadrupole 
memory device.

\begin{figure}[t]
\begin{center}
\vspace{-1mm}
\begin{tabular}{cc}
\includegraphics[width=0.18\textwidth]{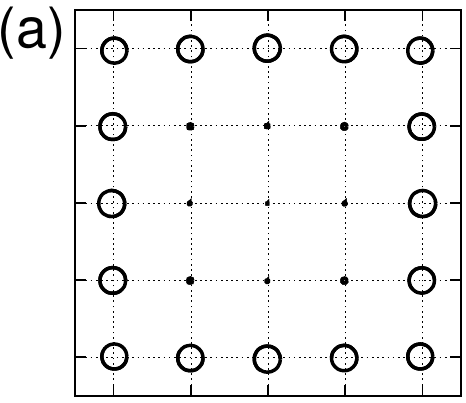} \hspace{1pt}
\includegraphics[width=0.18\textwidth]{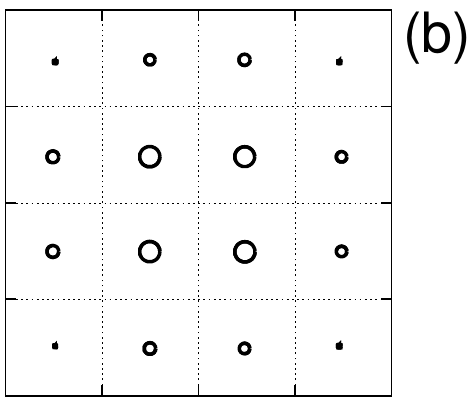} \\ 
\includegraphics[width=0.18\textwidth]{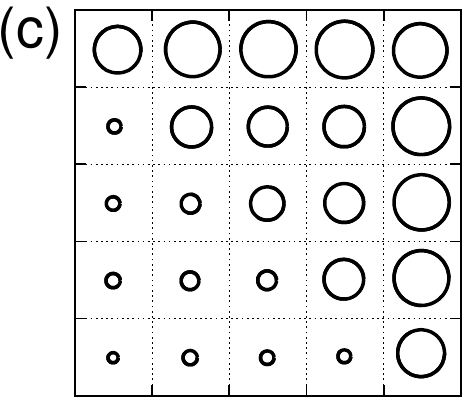} \hspace{1pt}
\includegraphics[width=0.18\textwidth]{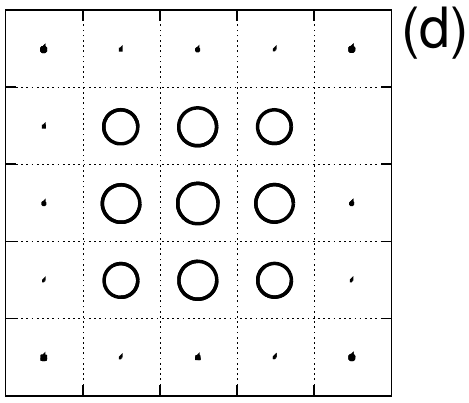}
\end{tabular}
\vspace{-1mm}
\caption{\label{zdep}{
  $z$ component of the Ti off-centering for the wires in Fig.~1.
  The largest circle represents a value of 15 pm.}}
\end{center}
\end{figure}

  Phase transformations were already proposed for vortex structures 
in nanodots~\cite{Naumov04}.
  In wires, as extended systems, the expression would be more appropriate
except for the fact that wires with sufficiently short-ranged interactions
do not sustain spontaneous symmetry breaking at any finite 
temperature~\cite{Mermin66}.
  Even assuming that the long range dipole-dipole interactions and the also
long-ranged strain mediated interactions do not invalidate the previous
argument, the discussions in this paper can be understood in terms of
the observed wires being of lengths smaller that the dipole-dipole
correlation length: any perturbation pins the order parameter
(the polarization across wires is experimentally observed after all). 
  As a final remark, the surface and edge control not only affects the 
polarization {\it across} the wires, but can also determine the conductance 
{\it along} the wire, since the deviations from (BaO)$_x$(TiO$_2$)$_y$ 
compositions give rise to doped wires, opening still further possibilities for
these systems, to be explored elsewhere.
  Evidently, different surface chemistries than the ones contemplated here
can be achieved depending on how the wires are produced. 
  The points remain, however, that interesting topologies are to be expected 
in thin ferroelectric wires, and that they are induced by the surfaces and 
edges.


  In summary, based on first-principles calculations, we predict ferroelectric 
patterns in BTO nanowires for different surface terminations, finding new 
textures that correspond to the topological winding numbers $n=0,+1,-1$ and 
$-3$ of point defects in 2D vector fields.
  A tendency of the $n=-3$ defect to decompose into one $+1$ and four $-1$
defects is observed, as well as the different mechanisms used by the field
to avoid discontinutities.

\begin{acknowledgments}
  J. W. Hong thanks China's Scholarship Council for its support.  
We acknowledge support through EPSRC and the computing resources of
Cambridge's CamGRID and the High Performance Computing Service.
\end{acknowledgments}


\end{document}